\documentclass[12pt]{article}

\newcommand{\beq}{\begin{equation}}
\newcommand{\eeq}{\end{equation}}
\newcommand{\beqa}{\begin{eqnarray}}
\newcommand{\eeqa}{\end{eqnarray}}
\newcommand{\beqar}{\begin{eqnarray*}}
\newcommand{\eeqar}{\end{eqnarray*}}

\newcommand{\al}{\alpha}
\newcommand{\be}{\beta}

\def\spa          {\ \ \ }
\def\non          {\nonumber}
\def\ha           {\mbox{$\frac{1}{2}$}}

\def\spa          {\ \ \ }
\def\mand         {\spa\mbox{and}\spa}

\def\Tr           {\mbox{\rm Tr}\,}
\def\STr          {\mbox{\rm STr}\,}
\def\Str          {\mbox{\rm Str}\,}

\def\cd           {{\cdot}}
\def\ran          {\rangle}
\def\lan          {\langle}
\def\fsC    {C\!\!\!\!/\,}
\def\fsH    {H\!\!\!\!/\,}
\newcommand{\del}{\delta}

\newcommand{\eps}{\epsilon}
\newcommand{\ga}{\gamma}
\newcommand{\Ga}{\Gamma}

\newcommand{\inn}{\!\cdot\!}

\newcommand{\lam}{\lambda}

\newcommand{\sig}{\sigma}

\newcommand{\z}{\zeta}

\newcommand{\labell}[1]{\label{#1}} 
\newcommand{\reef}[1]{(\ref{#1})}
\newcommand\prt{\partial}

\newcommand\bz{\bar{z}}

\def\sst#1{{\scriptscriptstyle #1}}
\def\0{{\sst{(0)}}}
\def\1{{\sst{(1)}}}
\def\2{{\sst{(2)}}}
\def\3{{\sst{(3)}}}
\def\4{{\sst{(4)}}}
\def\5{{\sst{(5)}}}
\def\6{{\sst{(6)}}}
\def\7{{\sst{(7)}}}
\def\8{{\sst{(8)}}}

\begin{document}
\baselineskip 18pt%
\begin{titlepage}
\vspace*{1mm}%
\hfill
\vbox{

    \halign{#\hfil         \cr
           } 
      }  
\vspace*{8mm}
\vspace*{8mm}%

\center{ {\bf \Large    On  Non-BPS 
Effective Actions of String Theory

}}\vspace*{3mm} \centerline{{\Large {\bf  }}}
\vspace*{5mm}
\begin{center}
{Ehsan Hatefi }

\vspace*{0.6cm}{
 Institute for Theoretical Physics, TU Wien
\quad\\
Wiedner Hauptstrasse 8-10/136, A-1040 Vienna, Austria 

\quad\\

Faculty of Physics, University of Warsaw, ul. Pasteura 5, 02-093 Warsaw, Poland
\quad\\

Mathematical Institute, Faculty of Mathematics,
Charles University, P-18675, CR}
\footnote{ehsan.hatefi@fuw.edu.pl, ehsanhatefi@gmail.com}

\vspace*{.3cm}
\end{center}
\begin{center}{\bf Abstract}\end{center}
\begin{quote}

We discuss some  physical prospective of the non-BPS effective actions  of type IIA and IIB superstring theories. By
dealing with the complete all three and four point functions, including  a closed string Ramond-Ramond (in terms of both its field strength and its potential), gauge (scalar) fields as well as a real tachyon and under 
symmetry structures, we find various restricted world volume and bulk  Bianchi identities.
 
  The complete forms of the non-BPS scattering amplitudes including their Chan-Paton factors are elaborated. All the singularity structures of the non-BPS amplitudes, their all orders $\alpha'$ higher derivative corrections, their contact terms and  various modified Bianchi identities are derived.

 Finally we show that scattering amplitudes computed in different super-ghost pictures are compatible when suitable Bianchi identities are imposed on the Ramond-Ramond fields. Moreover, we argue that the higher derivative expansion in powers of the momenta of the tachyon is universal.

 \end{quote}
\end{titlepage}
\section{Introduction}

D-branes have been realized to be the sources for Ramond-Ramond (RR) fields  \cite{Polchinski:1995mt,Witten:1995im}. RR couplings actually played important contributions to string theory. For instance to observe some of the application of RR couplings, one may consider the dissolving branes \cite{Douglas:1995bn}, K-theory and Myers effect \cite{Moore:97}.
The other applications to RR couplings are related to the $N^3$ phenomena for M5-branes, dS solutions, entropy growth  and geometrical applications to the effective actions \cite{Hatefi:2012bp}.
\vskip.1in

The spectrum of the so called non-BPS (unstable) branes includes massless, tachyon and an infinite number of massive states. There must be an Effective Field theory (EFT) for non-BPS branes where one integrates out all the massive states and hence the spectrum involves just the tachyon and massless states \cite{Sen:2004nf}. We will not point out cosmological applications for unstable branes. On general grounds, one might expect that D-branes and SD-branes have similar effective actions. The effective action of these branes has to have two parts.
It consists of the extensions of the usual DBI and Wess-Zumino (WZ) actions where the tachyon mode is embedded to these effective actions. 
By applying the Conformal Field Theory (CFT) methods \cite{Friedan:1985ge}, the leading order effective couplings of the fermions with tachyons  were found in \cite{Hatefi:2013mwa} as 
\beqa
S=-T_pV(T)\sqrt{-\det(\eta_{ab}+2\pi\alpha'F_{ab}-2\pi\alpha'\bar\Psi\ga_b\prt_a\Psi+\pi^2\alpha'^2\bar\Psi\ga^{\mu}\prt_a\Psi\bar\Psi\ga_{\mu}\prt_b\Psi + 2\pi\alpha' D_{a}T D_{b}T )}\nonumber\eeqa
In the above action, $F_{ab}$ is the field strength of the gauge field, $\bar\Psi\ga_{\mu}\prt^{\mu}\Psi$ is the kinetic term of fermion fields, $DT$ is the covariant derivative of the tachyon $(D_aT=\prt_a T-i[A_a,T])$. On the other hand the Chern-Simons action for BPS branes was constructed in \cite{li1996a}.
Using Boundary String Field Theory (BSFT), one talks about tachyon's kinetic term in DBI part \cite{Kraus:2000nj} as follows 
\beqa
S_{DBI}&\sim&\int d^{p+1}\sigma\, e^{-2\pi T^2} F(2\pi\alpha'D^aTD_aT),\qquad\qquad F(x)=\frac{4^xx\Gamma(x)^2}{2\Gamma(2x)}.
\eeqa
  The WZ action in BSFT approach is found to be
\beqa
S_{WZ}&=&\mu_p' \int_{\Sigma_{(p+1)}} C \wedge \Str e^{i2\pi\alpha'\cal F},\labell{WZ'}\eeqa 
 where $C_{(p+1)}$ is the RR potential $(p+1)$ form-field and super connection's curvature would be given by 
 \begin{displaymath}
i{\cal F} = \left(
\begin{array}{cc}
iF -\beta'^2 T^2 & \beta' DT \\
\beta' DT & iF -\beta'^2T^2 
\end{array}
\right),
\non\end{displaymath}
$\beta'$ is the normalisation constant and $\mu_p'$ is RR charge of the brane.
If we expand the exponential in \reef{WZ'}, then we obtain various couplings as follows 
\beqa
S_{WZ}=2\beta'\mu_p' (2\pi\alpha')\Tr\bigg(C_{p}\wedge DT + (2\pi\alpha')C_{p-2}\wedge DT\wedge F+
\frac{(2\pi\alpha')^2}{2}C_{p-4}\wedge F\wedge F\wedge DT\bigg)\label{esi88}\eeqa
For the sake of the higher derivative corrections, we work with the second approach of exploring effective actions, which is scattering amplitude formalism. In this approach tachyon's kinetic term is embedded into the DBI action as follows  
\beqa
S_{DBI}&\sim&\int
d^{p+1}\sigma \STr\left(\frac{}{}V({ T^iT^i})\sqrt{1+\frac{1}{2}[T^i,T^j][T^j,T^i])}\right.\labell{nonab1} \\
&&\qquad\qquad\left.
\times\sqrt{-\det(\eta_{ab}
+2\pi\alpha'F_{ab}+2\pi\alpha'D_a{ T^i}(Q^{-1})^{ij}D_b{ T^j})} \right)\,,\nonumber\eeqa  where $V({T^iT^i})=e^{-\pi{ T^iT^i}/2}$, 
$ Q^{ij}=I\delta^{ij}-i[T^i,T^j], T^1=T\sigma_1, T^2=T\sigma_2$ and $\sigma_1,\sigma_2$ are Pauli matrices.
 The trace in \reef{nonab1} 
should be   symmetric for all $F_{ab},D_a{ T^i}$, $[T^i,T^j]$ matrices. If all Chan-Paton factors are taken into account, then this action would produce consistent results with all momentum expansions of three and four point functions of a closed string RR and either two, three 
 tachyons or two tachyons two gauge/scalar fields  amplitude.
 
 \vskip.1in
 
On the stable point, tachyon potential  and its effective action get replaced to the known tachyon DBI action \cite{Sen:1999md,Bergshoeff:2000dq} with potential $T^4V(T^2)$. The WZ part of the action in this approach has the same formula as appeared in \reef{WZ'}. Using S-matrix method the normalisation constants of $\beta',\beta$ for non-BPS and brane anti brane system are discovered to be  $\beta'=\frac{1}{\pi}\sqrt{\frac{6\ln(2)}{\alpha'}}$ and $\beta=\frac{1}{\pi}\sqrt{\frac{2\ln(2)}{\alpha'}}$ \cite{Garousi:2007fk}. It is worth mentioning that  super conection's structure for WZ action was found by the S-matrix approach in \cite{Kennedy:1999nn}.
\vskip.1in

The aim of the paper is to show that scattering amplitudes computed in different super-ghost pictures are compatible when suitable Bianchi identities are imposed on the Ramond-Ramond (R-R) fields. Moreover, we argue that the higher-derivative expansion in powers of the momenta of the tachyon is universal.
\vskip.1in

The outline of this paper is as follows. First we find all three point functions including a gauge field, a tachyon and a closed string RR in all asymmetric and symmetric pictures of the closed string RR. By doing so, not only do we find some restricted Bianchi identities on both world volume and transverse directions of non-BPS branes but also explore all their infinite higher derivative corrections. It is believed that due to supersymmetry transformation BPS S-matrices do not generate Bianchi identity.  To get consistent results at the level of singularity structures for four point functions of the two gauge fields, a tachyon and a closed string RR in their all asymmetric and symmetric pictures, we discover various restricted Bianchi identities. Eventually we talk about a universal expansion for tachyon and start constructing all different singularity structures of $<V_{C^{-2}} V_{A^{0}} V_{A^{0}} V_{T^{0}}>$ as well as all order $\alpha'$  higher derivative corrections to various couplings of the type IIA, IIB superstring theories.
\section{All order $<V_{C^{-2}} V_{A^{0}} V_{T^{0}}>$  }

 In this section we would like to apply CFT methods to derive the complete S-matrix elements of a closed string RR, a gauge field and a tachyon in the world volume of a non-BPS brane. Total super ghost charge for disk level amplitude must be -2. First we choose an asymmetric closed string RR ( that carries total -2 super ghost charge) and hence the gauge field and a tachyon must be put in zero picture. This S-matrix can be achieved if one finds the correlation functions of the following vertex operators 
 \beqa
V_{T}^{(0)}(x) &=&  \alpha' ik_2\cd\psi(x) e^{\alpha' ik_2.\cd X(x)}\lam\otimes\sigma_1
\nonumber\\
V_{T}^{(-1)}(x) &=&e^{-\phi(x)} e^{\alpha' ik_2\cd X(x)}\lam\otimes\sigma_2\nonumber\\
V_A^{(-1)}(x)&=&e^{-\phi(x)}\xi_a\psi^a(x)e^{ \alpha'iq\inn X(x)}\lam\otimes \sigma_3 \nonumber\\
V_{A}^{(0)}(x) &=& \xi_{1a}(\partial^a X(x)+i\alpha'q.\psi\psi^a(x))e^{\alpha'iq.X(x)}\lam\otimes I\label{d4Vs}\\
V_{A}^{(-2)}(x) &=& e^{-2\phi(x)}V_{A}^{(0)}(x)\nonumber\\
V_{C}^{(-\frac{3}{2},-\frac{1}{2})}(z,\bar{z})&=&(P_{-}\fsC_{(n-1)}M_p)^{\alpha\beta}e^{-3\phi(z)/2}
S_{\al}(z)e^{i\frac{\alpha'}{2}p\cd X(z)}e^{-\phi(\bar{z})/2} S_{\be}(\bar{z})
e^{i\frac{\alpha'}{2}p\cd D \cd X(\bar{z})}\otimes\sigma_1\nonumber\\
V_{C}^{(-\frac{1}{2},-\frac{1}{2})}(z,\bar{z})&=&(P_{-}\fsH_{(n)}M_p)^{\alpha\beta}e^{-\phi(z)/2}
S_{\al}(z)e^{i\frac{\alpha'}{2}p\cd X(z)}e^{-\phi(\bar{z})/2} S_{\be}(\bar{z})
e^{i\frac{\alpha'}{2}p\cd D \cd X(\bar{z})}\otimes\sigma_3\sigma_1\nonumber
\eeqa

It is argued in  \cite{Sen:1999mg}
 that the vertices of a non-BPS D-brane need to carry internal degrees of freedom or Chan-Paton (CP) matrix. Because if we set the tachyon to zero, then the WZ effective action of non-BPS branes gets reduced to WZ action of BPS branes. Hence, we impose an identity internal CP matrix to all massless fields including gauge (scalar) and RR field in zero picture. It is also discussed in \cite{DeSmet:2000je} that Picture Changing Operator (PCO) carries CP matrix $\sig_3$.  It is also explained in  \cite{Hatefi:2013yxa} that the tachyon in zero and (-1) picture does carry $\sig_1$ and $\sig_2$ CP factor. We also know that the amplitude of a closed string RR and a tachyon $<V_{C^{-1}} V_{T^{-1}}>$ makes sense in the world volume of non-BPS branes. This fixes the CP factor of RR in $(-1/2, -1/2)$ picture for non-BPS branes to be $\sig_3\sig_1$. By applying PCO to RR in (-1) picture, we derive its CP factor in (-2) picture to be $\sig_1$ and the CP factor for the gauge field in (-1) picture to be $\sig_3$ where $\lambda$ is the external CP matrix for the U(N) gauge group.

We are looking for the disk level amplitude. The closed string will be located in the middle of disk whereas all open strings are located at the boundary of disk.  On-shell conditions are  
\beqa
  q^2=p^2=0, \quad  k_{2}^2=1/4  , q.\xi_1=0,
\nonumber\eeqa
The definitions of the RR's field strength and projection operator are
\begin{displaymath}
P_{-} =\ha (1-\ga^{11}), \quad
\fsH_{(n)} = \frac{a
_n}{n!}H_{\mu_{1}\ldots\mu_{n}}\ga^{\mu_{1}}\ldots
\ga^{\mu_{n}},
\non\end{displaymath}
For type IIA (IIB) $n=2,4$,$a_n=i$  ($n=1,3,5$,$a_n=1$) and spinor notation is   
\beqa
(P_{-}\fsH_{(n)})^{\al\be} =
C^{\al\del}(P_{-}\fsH_{(n)})_{\del}{}^{\be}.
\nonumber
\eeqa
We also apply the doubling trick so that all the holomorphic parts of the fields can be used. Thus the following change of variables works
\begin{displaymath}
\tilde{X}^{\mu}(\bar{z}) \rightarrow D^{\mu}_{\nu}X^{\nu}(\bar{z}) \ ,
\spa
\tilde{\psi}^{\mu}(\bar{z}) \rightarrow
D^{\mu}_{\nu}\psi^{\nu}(\bar{z}) \ ,
\spa
\tilde{\phi}(\bar{z}) \rightarrow \phi(\bar{z})\,, \mand
\tilde{S}_{\al}(\bar{z}) \rightarrow M_{\al}{}^{\be}{S}_{\be}(\bar{z})
 \ ,
\non\end{displaymath}

with 
\begin{displaymath}
D = \left( \begin{array}{cc}
-1_{9-p} & 0 \\
0 & 1_{p+1}
\end{array}
\right) \ ,\,\, \mand
M_p = \left\{\begin{array}{cc}\frac{\pm i}{(p+1)!}\ga^{i_{1}}\ga^{i_{2}}\ldots \ga^{i_{p+1}}
\eps_{i_{1}\ldots i_{p+1}}\,\,\,\,{\rm for\, p \,even}\\ \frac{\pm 1}{(p+1)!}\ga^{i_{1}}\ga^{i_{2}}\ldots \ga^{i_{p+1}}\ga_{11}
\eps_{i_{1}\ldots i_{p+1}} \,\,\,\,{\rm for\, p \,odd}\end{array}\right.
\non\end{displaymath}
\vskip .1in
Now one can use the following two-point functions for  $X^{\mu},\psi^\mu, \phi$, as below
\begin{eqnarray}
\lan X^{\mu}(z)X^{\nu}(w)\ran & = & -\frac{\alpha'}{2}\eta^{\mu\nu}\log(z-w) \ , \non \\
\lan \psi^{\mu}(z)\psi^{\nu}(w) \ran & = & -\frac{\alpha'}{2}\eta^{\mu\nu}(z-w)^{-1} \ ,\non \\
\lan\phi(z)\phi(w)\ran & = & -\log(z-w) \ .
\labell{prop2}\end{eqnarray}
The amplitude in asymmetric picture is given by
\beqa
&&\int dx_1 dx_2 dx_4 dx_5 (P_{-}\fsC_{(n-1)}M_p)^{\alpha\beta}(2 \alpha'ik_{2b} \xi_{1a}) (x_{45})^{-3/4} (I_1+I_2)\nonumber\\&&\times
|x_{12}|^{\alpha'^2k_1.k_2}|x_{14}x_{15}|^{\frac{\alpha'^2}{2}k_1.p} |x_{24}x_{25}|^{ \frac{\alpha'^2}{2}  k_2.p}|x_{45}|^{\frac{\alpha'^2}{4}p.D.p}\nonumber\eeqa
where $x_4=z=x+iy,x_5=\bar z=x-iy$ and 
\beqa
I_1&=&ik_{2}^{a}\bigg(\frac{x_{42}}{x_{12}x_{14}}+\frac{x_{52}}{x_{12}x_{15}}\bigg)
2^{-1/2}(x_{24}x_{25})^{-1/2}(x_{45})^{-3/4}(\gamma^{b} C^{-1})_{\alpha\beta}
\eeqa
One should make use of the Wick-like rule \cite{Liu:2003} to obtain the correlation function for $I_2$ 
\beqa
I_2 = 2ik_{1c}<:S_{\al}(x_4): S_{\be}(x_5):\psi^{c}\psi^{a}(x_1):\psi^{b}(x_2):> \nonumber\eeqa 
as follows
\beqa
 I_2 &=& \bigg((\Gamma^{bac} C^{-1})_{\alpha\beta}+\frac{2Re[x_{14}x_{25}]}{x_{12}x_{45}}\bigg(\eta^{bc}(\gamma^{a} C^{-1})_{\alpha\beta}-\eta^{ab}(\gamma^{c} C^{-1})_{\alpha\beta}\bigg)\bigg)
  \nonumber\\&&\times 2ik_{1c} 2^{-3/2}(x_{14}x_{15})^{-1}(x_{24}x_{25})^{-1/2}(x_{45})^{1/4}
  \nonumber\eeqa
It can be readily shown that the amplitude is  $SL(2,R)$ invariant. We use
 the gauge fixing as  $(x_1,x_2,z,\bar z)=(x,-x,i,-i)$ and the Jacobian is $J=-2i(1+x^2)$.
 After gauge fixing, one reveals that $I_1$ has zero contribution to the S-matrix. Because the integrand is an odd function while the the interval of the integral is symmetric.\footnote{ $\alpha'=2$ is set.} We introduce 
 $t = -\frac{\alpha'}{2}(k_1+k_2)^2$ and after gauge fixing $I_2$ is obtained by 
 \beqa
\int_{-\infty}^{\infty} dx (2x)^{-2t-1/2} 
(1+x^{2})^{-1/2 +2t} \bigg((\frac{1-x^2}{2ix})(\eta^{bc}\Tr
(P_{-}\fsC_{(n-1)}M_p\gamma^{a})\nonumber\\-\eta^{ab}\Tr
(P_{-}\fsC_{(n-1)}M_p\gamma^{c}))+\Tr
(P_{-}\fsC_{(n-1)}M_p\Gamma^{bac})\bigg)2^{3/2}k_{1c} k_{2b} \xi_{1a}   \nonumber\eeqa
 Note that the last two terms have just non-zero contribution to our amplitude.  The final answer for the amplitude is
\beqa
{\cal A}^{T^0,A^0,C^{-2}} &=&(\pi\beta'\mu_p')2\sqrt{\pi} \frac{\Ga[-t+1/4]}{\Ga[3/4-t]}
 \Tr
(P_{-}\fsC_{(n-1)}M_p\Gamma^{bac})k_{1c} k_{2b} \xi_{1a}\Tr(\lam_1\lam_2) \labell{amp383}\ .
\eeqa

To be able to match the leading order of the S-matrix with the following coupling in the EFT
 \beqa
    2i\beta'\mu'_p (2\pi\alpha')^2\int_{\Sigma_{p+1}} \Tr(C_{p-2}\wedge F\wedge DT),
    \label{jj}\eeqa
we use $ (\pi\beta'\mu_p'/2)$ as the normalization constant.
 $ \beta' $  and $  \mu_p' $ are known to be WZ normalisation  constant and RR brane's charge.
On the other hand, the result in symmetric cases (RR is written in (-1)-picture)  for ${\cal A}^{A^0,T^{-1},C^{-1}}$ can be derived as  
\beqa
{\cal A}^{A^0,T^{-1},C^{-1}}&=& 2ik_{1b} \xi_{1a}\Tr
(P_{-}\fsH_{(n)}M_p\Gamma^{ab})
\int_{-\infty}^{\infty} dx (2x)^{-2t-1/2} 
(1+x^{2})^{-1/2 +2t}  \label{jj12} \eeqa

Accordingly  ${\cal A}^{A^{-1},T^{0},C^{-1}}$ is found to be  
 \beqa
{\cal A}^{A^{-1},T^{0},C^{-1}}&=&2i k_{2b} \xi_{1a}\Tr
(P_{-}\fsH_{(n)}M_p\Gamma^{ba})
\int_{-\infty}^{\infty} dx (2x)^{-2t-1/2} 
(1+x^{2})^{-1/2 +2t}  \label{jj22}\eeqa
By applying momentum conservation along the world volume of brane $(k_1+k_2+p)^a=0$ and making comparisons between \reef{jj12} and \reef{jj22}, one clarifies that 
  the following Bianchi identities kept fixed in the presence of RR's field strength 
  \beqa
  p_b H_{a_0...a_{p-2}}\epsilon^{a_0...a_{p-2}ba}=p_a H_{a_0...a_{p-2}} \epsilon^{a_0...a_{p-2}ba}=0
  \label{jj33}\eeqa
 Note that all three point functions involving a closed string RR, a tachyon and a scalar field in all symmetric and asymmetric pictures of RR can also be computed.
 The result in symmetric picture for  ${\cal A}^{\phi^0,T^{-1},C^{-1}}$ is given by 
 \beqa
 4\xi_{1i} 
(P_{-}\fsH_{(n)}M_p)^{\alpha\beta}\bigg(k_{1a} (\Gamma^{ia} C^{-1})_{\alpha\beta}-p^i (C^{-1})_{\alpha\beta}\bigg)\int_{-\infty}^{\infty} dx (2x)^{-2t-1/2} 
(1+x^{2})^{-1/2 +2t}  \nonumber \eeqa
To get consistent result for the S-matrix,  in the presence of all different pictures of closed string RR, the restricted Bianchi identity \reef{jj33} must get replaced by the following Bianchi identity 
 \beqa
  p^i \epsilon^{a_0...a_{p}}H_{a_0...a_{p}}+p^a \epsilon^{a_0...a_{p-1}a}H^{i}_{a_0...a_{p-1}}=0
  \label{jj66}\eeqa
 This modified Bianchi identity works in the appearance of a scalar field and holds for all world volume and transverse directions of branes. The below trace is non-zero for $p+1= n+2$ case 
\beqa
\Tr\bigg(\fsC_{(n-1)}M_p (k_2.\ga)
(\xi.\ga)(k_1.\ga)\bigg)&=&\pm\frac{32}{(p-2)!}\eps^{a_{0}\cdots a_{p-3}bac}C_{a_{0}\cdots a_{p-3}}k_{1c} k_{2b} \xi_{1a}\nonumber\eeqa
The trace that has  $\gamma^{11}$ part, indicates that the following relation holds 
\beqa
  p>3 , H_n=*H_{10-n} , n\geq 5.
  \nonumber\eeqa
 Neither there are massless poles nor tachyon poles for this three point function. It is argued in \cite{Hatefi:2012wj} that the expansion of the non-BPS amplitudes in the presence of a closed string RR makes sense if one applies the following constraint 
\beqa
t=-p^ap_a\rightarrow \frac{-1}{4}.
\eeqa
Note that for the brane-anti brane configuration the above constraint gets replaced by $p^a p_a\rightarrow 0$ \cite{Hatefi:2012cp}. Hence, the precise momentum expansion for $CAT$ is  $t\rightarrow -1/4$. The expansion for Gamma function is  \beqa
\sqrt{\pi}\frac{\Ga[-t+1/4]}{\Ga[3/4-t]}
 &=&\pi \sum_{n=-1}^{\infty}c_n(t+1/4)^{n+1}
\ .\labell{taylor61}\nonumber
\eeqa
 with the following coefficients
\beqa
c_{-1}&=&1,c_0=2ln(2),c_1=\frac{1}{6}(\pi^2+12ln(2)^2),...\nonumber\eeqa
 
 An infinite number of the higher derivative corrections to a $C_{p-2}$, a tachyon and a gauge field can be found by producing the contact terms in an EFT and also by applying higher derivative corrections to the WZ  effective coupling  as follows
 
\beqa
\frac{2i\beta'\mu_p'}{(p-2)!}(2\pi\alpha')^2 C_{p-2}\wedge \Tr\left(\sum_{n=-1}^{\infty}c_n(\alpha')^{n+1}  D_{a_1}\cdots D_{a_{n+1}}F \wedge D^{a_1}...D^{a_{n+1}} DT\right) \labell{highaa}\eeqa
Let us deal with the  complete  $<V_{C^{-2}} V_{A^0} V_{A^0} V_{T^0}>$, to see what kinds of restricted Bianchi identities can be explored and also to see whether or not there are bulk singularity structures.
 \section{ The complete   $<V_{C^{-2}} V_{A^0} V_{A^0} V_{T^0}>$ amplitude}
 
In order to  find out the complete form of the scattering amplitude of a tachyon, a potential RR  $(p+1)$ form-field and two massless gauge fields  $<V_{C^{-2}} V_{A^0} V_{A^0} V_{T^0}>$ on the world volume of  non-BPS branes, one needs to employ all CFT techniques. To achieve all singularities and contact interactions, we make use of the vertex operators that are introduced earlier on. Note that as clarified in \cite{Hatefi1609} the CP factor of RR for brane anti brane system is different from the CP factor of non-BPS branes. RR vertex operators are introduced in \cite{Bianchi:1991eu}.  One might hint on some of the BPS and non-BPS scattering amplitudes in \cite{Barreiro:2013dpa}.

Recently $<V_{C^{-2}} V_{\phi^0} V_{\phi^0}V_{T^0}>$ analysis was done, however,  one cannot derive the result for  $<V_{C^{-2}} V_{A^0} V_{A^0} V_{T^0}>$ from it, by just exchanging the scalars fields to gauge fields. Because not all world volume couplings nor bulk terms have any effect in our new effective action. More crucially,  given the presence of the tachyonic string, we can not compare our effective actions with BPS branes' s effective action \cite{Hatefi:2015ora,Park:2008sg,Hatefi:2012ve}. Let us write down the compact and the closed form of the correlation functions \footnote{
$x_{ij}=x_i-x_j$, and $\alpha'=2$.}, whereas all the other kinematical relations can be found in  \cite{Hatefi1609}. 
\beqa
{\cal A'}^{C^{-2} A^0 A^0 T^0}&\sim& 2\int dx_{1}dx_{2}
dx_{3}dx_{4}dx_{5}(P_{-}\fsC_{(n-1)}M_p)^{\al\be}I \xi_{1a}\xi_{2b}x_{45}^{-3/4}\nonumber\\&&\times
\bigg((i\alpha'k_{3c}) a^c_1 \bigg[ a^a_1a^b_2 -\eta^{ab} x_{12}^{-2}\bigg]-\alpha'^2  k_{2d}k_{3c}a^a_1 a^{cbd}_{2}\nonumber\\&&-\alpha'^2  k_{1e}k_{3c}a^b_2
 a^{cae}_{3}-i{\alpha'}^3 k_{1e}k_{2d}k_{3c}
a_4^{cbdae}\bigg)\labell{ampe1},\eeqa
where
\beqa
I&=&|x_{12}|^{\alpha'^2 k_1.k_2}|x_{13}|^{\alpha'^2 k_1.k_3}|x_{14}x_{15}|^{\frac{\alpha'^2}{2} k_1.p}|x_{23}|^{\alpha'^2 k_2.k_3}|
x_{24}x_{25}|^{\frac{\alpha'^2}{2} k_2.p}
|x_{34}x_{35}|^{\frac{\alpha'^2}{2} k_3.p}|x_{45}|^{\frac{\alpha'^2}{4}p.D.p},\nonumber\\
a^a_1&=&ik_{2}^{a}\bigg(\frac{x_{42}}{x_{14}x_{12}}+\frac{x_{52}}{x_{15}x_{12}}\bigg)+ik_{3}^{a}\bigg(\frac{x_{43}}{x_{14}x_{13}}+\frac{x_{53}}{x_{15}x_{13}}\bigg),\nonumber\\
a^b_2&=&-ik_{1}^{b}\bigg(\frac{x_{14}}{x_{42}x_{12}}+\frac{x_{15}}{x_{52}x_{12}}\bigg)-ik_{3}^{b}\bigg(\frac{x_{43}}{x_{42}x_{23}}+\frac{x_{53}}{x_{52}x_{23}}\bigg),\nonumber\\
a^{c}_{1}&=&2^{-1/2}x_{45}^{-3/4}(x_{34}x_{35})^{-1/2} (\gamma^{c}C^{-1})_{\alpha\beta} ,\nonumber\\
a^{cbd}_{2}&=&2^{-3/2}x_{45}^{1/4}(x_{34}x_{35})^{-1/2}(x_{24}x_{25})^{-1}\bigg\{(\Gamma^{cbd}C^{-1})_{\alpha\beta}+\alpha' h_1
\frac{Re[x_{24}x_{35}]}{x_{23}x_{45}}\bigg\}
,\nonumber\\
a^{cae}_{3}&=&2^{-3/2}x_{45}^{1/4}(x_{34}x_{35})^{-1/2}(x_{14}x_{15})^{-1}\bigg\{(\Gamma^{cae}C^{-1})_{\alpha\beta}+\alpha' h_2
\frac{Re[x_{14}x_{35}]}{x_{13}x_{45}}\bigg\},\nonumber\\
h_1&=&\eta^{dc} (\gamma^{b}C^{-1})_{\alpha\beta}-\eta^{bc} (\gamma^{d}C^{-1})_{\alpha\beta},\nonumber\\
h_2&=&\eta^{ec} (\gamma^{a}C^{-1})_{\alpha\beta}-\eta^{ac} (\gamma^{e}C^{-1})_{\alpha\beta}.\nonumber\eeqa
The last fermionic correlator that has just world volume indices $a_4^{cbdae}=<:S_{\al}(x_4): S_{\be}(x_5):\psi^{e}\psi^{a}(x_1):\psi^{d}\psi^{b}(x_2):\psi^{c}(x_3):>$ can be explored by considering all possible interactions as follows:
\beqa
a_4^{cbdae}&=&
\bigg\{(\Gamma^{cbdae}C^{-1})_{{\alpha\beta}}+\alpha' h_3\frac{Re[x_{14}x_{25}]}{x_{12}x_{45}}+
\alpha' h_4\frac{Re[x_{14}x_{35}]}{x_{13}x_{45}}+\alpha' h_5 \frac{Re[x_{24}x_{35}]}{x_{23}x_{45}}\nonumber\\&&+ \alpha'^2h_6\bigg(\frac{Re[x_{14}x_{35}]}{x_{13}x_{45}}\bigg)\bigg(\frac{Re[x_{14}x_{25}]}{x_{12}x_{45}}\bigg)
+\alpha'^2 h_7
\bigg(\frac{Re[x_{14}x_{25}]}{x_{12}x_{45}}\bigg)^{2}\nonumber\\&&
+\alpha'^2 h_8 \bigg(\frac{Re[x_{14}x_{25}]}{x_{12}x_{45}}\bigg)\bigg(\frac{Re[x_{24}x_{35}]}{x_{23}x_{45}}\bigg)
\bigg\}2^{-5/2}x_{45}^{5/4}(x_{14}x_{15}x_{24}x_{25})^{-1}(x_{34}x_{35})^{-1/2},\nonumber\\
h_3&=&\bigg(\eta^{ed}(\Gamma^{cba}C^{-1})_{\alpha\beta}
-\eta^{eb}(\Gamma^{cda}C^{-1})_{\alpha\beta}-
\eta^{ad}(\Gamma^{cbe}C^{-1})_{\alpha\beta}
+\eta^{ab}(\Gamma^{cde}C^{-1})_{\alpha\beta}
\bigg),\nonumber\\
h_4&=&\bigg(\eta^{ec}(\Gamma^{bda}C^{-1})_{\alpha\beta}
-\eta^{ac}(\Gamma^{bde}C^{-1})_{\alpha\beta}
\bigg),\nonumber\\
h_5&=&\bigg(\eta^{dc}(\Gamma^{bae}C^{-1})_{\alpha\beta}
-\eta^{bc}(\Gamma^{dae}C^{-1})_{\alpha\beta}\bigg),\nonumber\\
h_6&=&\bigg(\eta^{ed}\eta^{ac}(\gamma^{b}C^{-1})_{\alpha\beta}- \eta^{eb}\eta^{ac}(\gamma^{d}C^{-1})_{\alpha\beta}- \eta^{ec}\eta^{ad}(\gamma^{b}C^{-1})_{\alpha\beta}+ \eta^{ec}\eta^{ab}(\gamma^{d}C^{-1})_{\alpha\beta}\bigg),\nonumber\\
h_7&=&\bigg(- \eta^{ed}\eta^{ab}(\gamma^{c}C^{-1})_{\alpha\beta}+ \eta^{eb}\eta^{ad}(\gamma^{c}C^{-1})_{\alpha\beta}\bigg),\label{hh22}\\
h_8&=&\bigg(-\eta^{ed}\eta^{bc}(\gamma^{a}C^{-1})_{\alpha\beta}+\eta^{eb}\eta^{dc}(\gamma^{a}C^{-1})_{\alpha\beta}
+\eta^{ad}\eta^{bc}(\gamma^{e}C^{-1})_{\alpha\beta}
- \eta^{ab}\eta^{dc}(\gamma^{e}C^{-1})_{\alpha\beta}\bigg).\nonumber\eeqa
We wrote all the S-matrix elements so that  SL(2,R) invariance can be manifestly shown. By fixing three position of vertices, we can get rid of  the volume of killing group. In order to get the algebraic answer for the amplitude, we fix the positions of open strings as
\beqar
 x_{1}&=&0 ,\qquad x_{2}=1,\qquad x_{3}\rightarrow \infty,
 \eeqar 
Eventually one needs to take a 2D complex integrals on the location of the closed string RR on the upper half plane as below
\beqa 
 \int d^2 \!z |1-z|^{a} |z|^{b} (z - \bar{z})^{c}
(z + \bar{z})^{d} 
 \eeqa
where  $d=0,1,2$ and  $a,b,c$ are written down in terms of the following Mandelstam variables:
 \beqa
s&=&\frac{-\alpha'}{2}(k_1+k_3)^2, t=\frac{-\alpha'}{2}(k_1+k_2)^2, u=\frac{-\alpha'}{2}(k_2+k_3)^2, s'=s+\frac{1}{4},u'=u+\frac{1}{4}.
\nonumber\eeqa
For $d=0,1$ and $d=2,3$ the algebraic solutions for integrals are gained in  \cite{Fotopoulos:2001pt},\cite{Hatefi:2016wof} accordingly.
We just demonstrate the final form of the amplitude as follows

\beqa {\cal A'}^{C^{-2} A^0 A^{0} T^0}&=&{\cal A'}_{1}+{\cal A'}_{2}+{\cal A'}_{3}\labell{44aa}\eeqa 
where 
\beqa
{\cal A'}_{1}&=&-2^{3/2}i\xi_{1a}\xi_{2b}k_{1e}k_{3c}k_{2d}
\Tr(P_{-}\fsC_{(n-1)}M_p\Gamma^{cbdae})(t+s'+u')L_3 
\nonumber\\
{\cal A'}_{2}&=&2^{3/2}iL_1\bigg((k_{1c}+k_{2c}+k_{3c})(-t\xi_{1a}\xi_{2b}\Tr(P_{-}\fsC_{(n-1)}M_p\Gamma^{cba}))
+2\Tr(P_{-}\fsC_{(n-1)}M_p \Gamma^{cde})
\nonumber\\&&\times k_{3c}k_{2d}k_{1e}\xi_1.\xi_2\bigg)+2^{3/2}i
\bigg(\xi_{2b}\left(-2k_3.\xi_1k_{2d}L_2(k_{3c}+k_{1c})+2k_2.\xi_1k_{3c} k_{2d}L_1\right)\nonumber\\&&\times\Tr(P_{-}\fsC_{(n-1)}M_p \Gamma^{dbc})
-2k_2.\xi_1k_{3c}k_{1e}\xi_{2b}
\Tr(P_{-}\fsC_{(n-1)}M_p\Gamma^{cbe})L_1
-2\frac{s'}{u'}k_3.\xi_2k_{1e}\xi_{1a}\nonumber\\&&\times(k_{3c}+k_{2c})
\Tr(P_{-}\fsC_{(n-1)}M_p\Gamma^{cae})L_2
-2k_1.\xi_2k_{3c}k_{2d}\xi_{1a}
\Tr(P_{-}\fsC_{(n-1)}M_p\Gamma^{cda})L_1\nonumber\\&&
-2k_1.\xi_2k_{3c}k_{1e}\xi_{1a}
\Tr(P_{-}\fsC_{(n-1)}M_p\Gamma^{cae})L_1
\bigg)
\nonumber\\
{\cal A'}_{3}&=&2^{3/2}i 
\Tr(P_{-}\fsC_{(n-1)}M_p\gamma^{c})L_3(k_{1c}+k_{2c}+k_{3c})\bigg[t(k_3.\xi_1)(k_3.\xi_2)
-\frac{1}{2}(\xi_1.\xi_2)u's'\nonumber\\&&
-(k_3.\xi_1)(k_1.\xi_2)u'
-s'(k_3.\xi_2)(k_2.\xi_1)\bigg]
\labell{11bb}\eeqa
  The functions  
 $L_1,L_2,L_3$ are 
 \beqa 
L_1&=&(2)^{-2(t+s+u)-1}\pi{\frac{\Gamma(-u+\frac{3}{4})
\Gamma(-s+\frac{3}{4})\Gamma(-t)\Gamma(-t-s-u)}
{\Gamma(-u-t+\frac{3}{4})\Gamma(-t-s+\frac{3}{4})\Gamma(-s-u+\frac{1}{2})}}\nonumber\\
L_2&=&(2)^{-2(t+s+u)-1}\pi{\frac{\Gamma(-u+\frac{3}{4})
\Gamma(-s-\frac{1}{4})\Gamma(-t+1)\Gamma(-t-s-u)}
{\Gamma(-u-t+\frac{3}{4})\Gamma(-t-s+\frac{3}{4})\Gamma(-s-u+\frac{1}{2})}}\nonumber\\
L_3&=&(2)^{-2(t+s+u)}\pi{\frac{\Gamma(-u+\frac{1}{4})
\Gamma(-s+\frac{1}{4})\Gamma(-t+\frac{1}{2})\Gamma(-t-s-u-\frac{1}{2})}{\Gamma(-u-t+\frac{3}{4})
\Gamma(-t-s+\frac{3}{4})\Gamma(-s-u+\frac{1}{2})}}\nonumber\eeqa
 
 This amplitude satisfies Ward identities related to both gauge fields. We expand the amplitude in such a way that all tachyon and massless poles can be obtained from the EFT. Finally we start producing all contact interactions to all orders in $\alpha'$ as well.
One may think that the amplitude in asymmetric case  has also non-zero terms for $p+1=n$ case, however, note that these terms that are not gauge invariant, will be disappeared from the final form of the S-matrix. These terms are \beqa
2^{1/2}i \Tr(P_{-}\fsC_{(n-1)}M_p\gamma^{c})\bigg[
 2(k_2.\xi_1)(k_1.\xi_2) k_{3c} (-4M_5+4K_2+M_9-4(\frac{1}{4}M_9+K_2-M_5))\nonumber\\
 +\xi_{2c}k_2.\xi_1\bigg(4u'(-M_4+\frac{1}{2}M_5+\frac{1}{2}M_{11}-\frac{1}{4}M_9)-4s'(\frac{1}{4}M_9-\frac{1}{2}M_{5})\bigg)\nonumber\\
 +\xi_{2c}k_3.\xi_1\bigg(-2u'(-M_{11}+\frac{1}{2}M_9)+4t(\frac{1}{4}M_9-\frac{1}{2}M_{5})\bigg)\nonumber\\
 +\xi_{1c}k_3.\xi_2\bigg(2s'M_{11}-s'M_9-4t(\frac{1}{4}M_9-\frac{1}{2}M_{11}+M_4-\frac{1}{2}M_{5})\bigg)\nonumber\\
 +\xi_{1c}k_1.\xi_2\bigg(-2s'M_{5}+s'M_9+4u'(\frac{1}{4}M_9-\frac{1}{2}M_{11}+M_4-\frac{1}{2}M_{5})\bigg)
\bigg]
\label{nn11}\eeqa
At the end of the day the sum of all coefficients of each parenthesis of \reef{nn11} is zero. This means that they disappear from the ultimate form of the  world volume amplitudes. All  $K_2,  M$ functions are written in terms of Gamma functions and for the sake of this paper we  will not mention their explicit forms. This also confirms that there is no bulk singularity term for this S-matrix. In the next section we would like to find the symmetric analysis of the same S-matrix.
\section{The complete   $<V_{C^{-1}} V_{A^{0}} V_{A^{-1}} V_{T^{0}}>$ amplitude}

This S-matrix in terms of the field strength of the closed string RR, that is, $<V_{C^{-1}}  V_{A^0} V_{A^{-1}}V_{T^0}>$  has not been calculated yet. Making use of the CFT, we try to explore the complete amplitude of $<V_{C^{-1}}  V_{A^0} V_{A^{-1}}V_{T^0}>$  on the  world volume of non-BPS branes as well. It  can be found by the following correlations
\beqa
{\cal A''}^{C^{-1} A^0 A^{-1} T^0}&=&-2\int dx_{1}dx_{2} dx_{3}dx_{4}dx_{5}(P_{-}\fsH_{(n)}M_p)^{\al\be}I\xi_{1a}\xi_{2b}x_{45}^{-1/4}(x_{24}x_{25})^{-1/2}\nonumber\\&&\times
\bigg(i\alpha' k_{3c} a^a_1a^{cb}_2-\alpha'^2 k_{1d}k_{3c}I_2^{cbad}\bigg)\labell{amp3}\eeqa
\beqa 
I&=&|x_{12}|^{4k_1.k_2}|x_{13}|^{4k_1.k_3}|x_{14}x_{15}|^{2k_1.p}|x_{23}|^{4k_2.k_3}|x_{24}x_{25}|^{2k_2.p}
|x_{34}x_{35}|^{2k_3.p}|x_{45}|^{p.D.p}\nonumber\\
a^a_1&=&ik_2^{a}\bigg(\frac{x_{42}}{x_{14}x_{12}}+\frac{x_{52}}{x_{15}x_{12}}\bigg)
+ik_3^{a}\bigg(\frac{x_{43}}{x_{14}x_{13}}+\frac{x_{53}}{x_{15}x_{13}}\bigg)\label{rr33}\eeqa
One needs to know the following correlation functions  
\beqa
a^{cb}_2&=&<:S_{\al}(x_4): S_{\be}(x_5):\psi^{b}(x_2):\psi^{c}(x_3):>=\bigg\{(\Gamma^{cb}C^{-1})_{\alpha\beta}-2\eta^{bc}
\frac{Re[x_{24}x_{35}]}{x_{23}x_{45}}\bigg\}\nonumber\\&&\times
2^{-1}(x_{24}x_{25}x_{34}x_{35})^{-1/2}x_{45}^{-1/4}\eeqa

One also needs to obtain the correlation function of a current and two fermion fields in two different locations in the presence of two spin operator, that is, 
$I_2^{cbad}=<:S_{\al}(x_4): S_{\be}(x_5):\psi^{d}\psi^{a}(x_1):\psi^{b}(x_2):\psi^{c}(x_3):>$ as
\beqa
I_2^{cbad}&=&
\bigg\{(\Gamma^{cbad}C^{-1})_{{\alpha\beta}}+\alpha' b_1\frac{Re[x_{14}x_{25}]}{x_{12}x_{45}}+
\alpha' b_2\frac{Re[x_{24}x_{35}]}{x_{23}x_{45}}+\alpha' b_3 \frac{Re[x_{14}x_{35}]}{x_{13}x_{45}}\nonumber\\&&+ \alpha'^2b_4\bigg(\frac{Re[x_{14}x_{35}]}{x_{13}x_{45}}\bigg)\bigg(\frac{Re[x_{14}x_{25}]}{x_{12}x_{45}}\bigg)
\bigg\}2^{-2}x_{45}^{3/4}(x_{34}x_{35}x_{24}x_{25})^{-1/2}(x_{14}x_{15})^{-1},\nonumber\\
b_1&=&\bigg(\eta^{bd}(\Gamma^{ca}C^{-1})_{\alpha\beta}
-\eta^{ab}(\Gamma^{cd}C^{-1})_{\alpha\beta}
\bigg),\nonumber\\
b_2&=&\bigg(-\eta^{bc}(\Gamma^{ad}C^{-1})_{\alpha\beta}
\bigg),\nonumber\\
b_3&=&\bigg(-\eta^{cd}(\Gamma^{ba}C^{-1})_{\alpha\beta}
+\eta^{ac}(\Gamma^{bd}C^{-1})_{\alpha\beta}\bigg),\nonumber\\
b_4&=&\bigg(-\eta^{ac}\eta^{db}+\eta^{ab}\eta^{dc}\bigg)(C^{-1})_{\alpha\beta},\nonumber\eeqa

 We fixed three positions of the open strings as $ x_{1}=0 ,x_{2}=1,x_{3}\rightarrow \infty$,and one essentially takes integration on the position of closed string RR which is  a 2D complex integrals on upper half plane.  Having set the gauge fixing, one would find the complete form of the integrand for  ${\cal A}^{C^{-1} A^0 A^{-1} T^0}$ as follows:
\beqa 
&&-2\xi_{1a}\xi_{2b} (P_{-}\fsH_{(n)}M_p)^{\alpha\beta}\int d^2 \!z |1-z|^{2t+2u-3/2} |z|^{2t+2s+1/2} (z - \bar{z})^{-2(t+s+u)-1}
\nonumber\\&&\times\bigg[-k_{3c}(2k_{2a}-|z|^{-2}(k_{2a}+k_{3a})(z+\bz) )((\Gamma^{cb}C^{-1})_{\alpha\beta}+2\eta^{bc}(\frac{1-x}{x_{45}}))
\nonumber\\&&-k_{1d}k_{3c}|z|^{-2}x_{45}\bigg((\Gamma^{cbad}C^{-1})_{\alpha\beta}+\frac{2x}{x_{45}}
(\eta^{bd}(\Gamma^{ca}C^{-1})_{\alpha\beta}-\eta^{ab}(\Gamma^{cd}C^{-1})_{\alpha\beta}-\eta^{bc}(\Gamma^{ad}C^{-1})_{\alpha\beta}
\nonumber\\&&-\eta^{cd}(\Gamma^{ba}C^{-1})_{\alpha\beta}+\eta^{ac}(\Gamma^{bd}C^{-1})_{\alpha\beta})-\frac{2|z|^{2}}{x_{45}}
(\eta^{bd}(\Gamma^{ca}C^{-1})_{\alpha\beta}-\eta^{ab}(\Gamma^{cd}C^{-1})_{\alpha\beta})\nonumber\\&&+\frac{2}{x_{45}}\eta^{bc}(\Gamma^{ad}C^{-1})_{\alpha\beta}+\frac{x^2-x|z|^{2}}{x_{45}^2}(-4\eta^{ac}\eta^{db}+4\eta^{ab}\eta^{dc})C^{-1}_{\alpha\beta}\bigg)
\bigg] 
 \eeqa
The final answer is given by
\beqa {\cal A''}^{C^{-1} A^0 A^{-1} T^0}&=&{\cal A''}_{1}+{\cal A''}_{2}+{\cal A''}_{3}\labell{44ac1}\eeqa 
where 
\beqa
{\cal A''}_{1}&=&2\xi_{1a}\xi_{2b}k_{1d}k_{3c}
\Tr(P_{-}\fsH_{(n)}M_p\Gamma^{cbad}
)(t+s'+u')L_3 
\nonumber\\
{\cal A''}_{2}&=&-2\bigg(\xi_{2b}\left(-2k_2.\xi_1k_{3c} L_1+2k_3.\xi_1L_2(k_{3c}+k_{1c})\right)\Tr(P_{-}\fsH_{(n)}M_p \Gamma^{cb})\nonumber\\&&
+2k_1.\xi_2k_{3c}\xi_{1a}
\Tr(P_{-}\fsH_{(n)}M_p\Gamma^{ca})L_1
-2\frac{s'}{u'}k_3.\xi_2k_{1d}\xi_{1a}
\Tr(P_{-}\fsH_{(n)}M_p\Gamma^{ad})L_2\bigg)
\nonumber\\&&
+2L_1\left(t\xi_{1a}\xi_{2b}\Tr(P_{-}\fsH_{(n)}M_p\Gamma^{ba})
+2k_{3c}k_{1d}\xi_1.\xi_2\Tr(P_{-}\fsH_{(n)}M_p \Gamma^{cd})\right)
\nonumber\\
{\cal A''}_{3}&=&2\Tr(P_{-}\fsH_{(n)}M_p)L_3 \bigg[-t(k_3.\xi_1)(k_3.\xi_2)+(k_3.\xi_2)(k_2.\xi_1)s'
\nonumber\\&&+(k_3.\xi_1)(k_1.\xi_2)u'+\frac{1}{2}(\xi_1.\xi_2)u's'\bigg]
\labell{113a}\eeqa

On the other hand, this amplitude for the following picture  $<V_{C^{-1}}V_{A^0} V_{A^0} V_{T^{-1}} >$ was computed to be 
\beqa {\cal A}^{C^{-1} A^0 A^0 T^{-1}}&=&{\cal A}_{1}+{\cal A}_{2}+{\cal A}_{3}\labell{44}\eeqa 
where 
\beqa
{\cal A}_{1}&=&-2i\xi_{1a}\xi_{2b}k_{1e}k_{2d}
\Tr(P_{-}\fsH_{(n)}M_p\Gamma^{bdae}
)(t+s'+u')L_3 
\nonumber\\
{\cal A}_{2}&=&2\bigg\{\bigg[k_{2d}\xi_{2b}\left(-2k_2.\xi_1 L_1+2k_3.\xi_1L_2\right)\Tr(P_{-}\fsH_{(n)}M_p \Gamma^{db})\nonumber\\&&
-2k_1.\xi_2k_{2d}\xi_{1a}
\Tr(P_{-}\fsH_{(n)}M_p\Gamma^{ad})L_1\bigg]+\bigg[k_{1d}\xi_{1a}\left(2k_1.\xi_2 L_1-2k_3.\xi_2\frac{s'}{u'}L_2\right)
\nonumber\\&&\times\Tr(P_{-}\fsH_{(n)}M_p \Gamma^{da})
+2k_2.\xi_1k_{1d}\xi_{2b}
\Tr(P_{-}\fsH_{(n)}M_p\Gamma^{bd})L_1\bigg]\nonumber\\
&&-L_1\left(-t\xi_{1a}\xi_{2b}\Tr(P_{-}\fsH_{(n)}M_p\Gamma^{ba})
+2k_{2d}k_{1e}\xi_1.\xi_2\Tr(P_{-}\fsH_{(n)}M_p \Gamma^{de}
)\right)\bigg\}
\nonumber\\
{\cal A}_{3}&=&\bigg[
-t(k_3.\xi_1)(k_3.\xi_2)+(k_3.\xi_2)(k_2.\xi_1)s'+(k_3.\xi_1)(k_1.\xi_2)u'+\frac{1}{2}(\xi_1.\xi_2)u's'\bigg]\nonumber\\&&\times-2i\Tr(P_{-}\fsH_{(n)}M_p)L_3
\labell{111}\eeqa
In the next section we address tachyon's momentum expansion to be able to expand our complete S-matrix and finally we start generating its non-zero couplings.
\section{Tachyon's momentum expansion}

In  \cite{Hatefi:2012wj}   it is conjectured that the momentum expansion for tachyon is universal. Given the momentum conservation on the world volume of brane 
for  a two point function of a closed string RR and a tachyon, one reveals that $k_a+p_a=0$, therefore  $p^ap_a$  must be sent to mass of the tachyon ($k^2=-m^2$). Hence, it is easy to understand the fact that
\beqa
p^ap_a \rightarrow  \frac{1}{4}\nonumber\eeqa
and that is just possible for SDbranes or euclidean branes. This means that  amplitude makes sense for non-BPS SD-branes \cite{Gutperle:2002ai}.  
From the kinetic term of the tachyons in DBI action, one reveals that  the coupling of the two tachyons and a gauge field is non zero, so to be able to produce all tachyon and massless poles of the EFT, we need to employ a unique expansion for all non-BPS branes. Two Mandelstam variables should be sent to the mass of the tachyon as follows   
\beqa
 s+t+u=-p^a p_a-\frac{1}{4}, \quad t\rightarrow 0, s\rightarrow -\frac{1}{4}, u\rightarrow -\frac{1}{4}\label{ess3}\eeqa
  $(i\mu'_p\beta'\pi^{1/2})$ is  the normalisation constant and the closed form of the expansions are
\beqa
L_1&=&-\pi^{3/2}\bigg(\frac{1}{t}\sum_{n=-1}^{\infty}b_n(u'+s')^{n+1}+\sum_{p,n,m=0}^{\infty}e_{p,n,m}t^{p}(s'u')^{n}(s'+u')^m\bigg)\nonumber\\
L_2&=&-\pi^{3/2}\bigg(\frac{1}{s'}\sum_{n=-1}^{\infty}b_n(u'+t)^{n+1}+\sum_{p, n,m=0}^{\infty}e_{p,n,m}s'^{p}(tu')^{n}(t+u')^m\bigg)\labell{high52}\eeqa
some of the above coefficients are found.\footnote{\beqa 
L_3&=&-{\pi^{5/2}} \sum_{p,n,m=0}^{\infty}\left(c_n(s'+t+u')^n+c_{n,m}\frac{s'^nu'^m +s'^mu'^n}{(t+s'+u')}+f_{p,n,m}(s'+t+u')^p (s'+u')^n(s'u')^{m}\right)
\nonumber\\
&&b_{-1}=1,\,b_0=0,\,b_1=\frac{1}{6}\pi^2 ,e_{1,0,0}=\frac{1}{6}\pi^2 c_0=0,c_1=\frac{\pi^2}{3},c_{1,0}=c_{0,1}=0, f_{0,0,1}=4\z(3) \nonumber
\eeqa}
Having taken \reef{high52}, we would understand that  $L_1,L_2,L_3$  have $t$-channel gauge fields, $s', u'$-channel tachyonic singularities and another kind of  $(s'+t+u')$-channel tachyon poles accordingly. In the next section we first try to make comparisons on singularity structures as well as all contact terms.  We then reconstruct all singularities in EFT and get to derive the restricted world volume Bianchi identities for non-BPS branes.

\section{Singularities and restricted Bianchi identities  }

Let us first compare singularity structures between $<V_{C^{-2}} V_{A^0} V_{A^0}V_{T^{0}}> $  and
  $<V_{C^{-1}} V_{A^0} V_{A^{-1}}V_{T^0}>$ with  \reef{111}.
If we use  momentum conservation   $k_{1a}+k_{2a}+k_{3a}=-p_a$  and $pC=H$ to the complete ${\cal A'}_{3}$ of 
 \reef{44aa}, then we are able to precisely produce all $(s'+t+u')$ channel poles  ${\cal A}_{3}$  of  \reef{111}.
 The first term ${\cal A'}_{2}$ exactly produces the 7th term ${\cal A}_{2}$  that has  tachyon singularities. Replacing $k_{3c}=-(k_{1c}+k_{2c}+p_c)$ to  the second term of ${\cal A'}_{2}$, we gain
 \beqa
 - 2^{5/2}iL_1 \Tr(P_{-}\fsC_{(n-1)}M_p \Gamma^{cde}) (k_{1c}+k_{2c}+p_c)k_{2d}k_{1e}\xi_1.\xi_2\label{5672}\eeqa
  \reef{5672}  is symmetric under both $k_{1c},k_{2c}$  and also is antisymmetric in terms $\epsilon^{a_0...a_{p-3}cde}$ inside the trace, therefore  $k_{1c},k_{2c}$ have no contribution to  our coupling. Using $pC=H$ we derive the 8th term of ${\cal A}_{2}$ that has  $(s'+t+u')$ tachyon singularities. Above arguments are also held for the third term of  ${\cal A'}_{2}$ , namely using momentum conservation, the third term of  ${\cal A'}_{2}$  reconstructs the second term of ${\cal A}_{2}$ that has $s'$-channel tachyon poles. If we apply momentum conservation to the 6th term of ${\cal A'}_{2}$, we find the following interaction 
 \beqa
  2^{5/2}i\frac{s'}{u'}  L_2 k_3.\xi_2 k_{1e} \xi_{1a} \Tr(P_{-}\fsC_{(n-1)}M_p \Gamma^{cae}) (k_{1c}+p_c)
\label{56713}\eeqa
 
  $k_{1c}$ has no effect to the above interaction and using $pC=H$, \reef{56713} regenerates exactly all $u'$ channel tachyon singularities (the 5th term of  ${\cal A}_{2}$). Having applied momentum conservation to the 4th term  ${\cal A'}_{2}$ we  obtain
 \beqa
  2^{3/2}i L_1 2k_2.\xi_1 k_{2d} \xi_{2b} \Tr(P_{-}\fsC_{(n-1)}M_p \Gamma^{cbd}) (k_{1c}+k_{2c}+p_c)
\label{567u}\eeqa
  $k_{2c}$ has no contribution to the above interaction and using $pC=H$, one reproduces exactly the first term of  ${\cal A}_{2}$. Now adding the contribution $k_{1c}$ of \reef{567u} with the 5th term of  ${\cal A'}_{2}$ we obtain
\beqa
  2^{3/2}i L_1 2k_2.\xi_1 k_{1c} \xi_{2b} \Tr(P_{-}\fsC_{(n-1)}M_p \Gamma^{cbd}) (k_{2d}+k_{3d})
\label{5674}\eeqa
which is precisely  the 6th term of ${\cal A}_{2}$. By applying momentum conservation to the 7th term ${\cal A'}_{2}$ we get
\beqa
 2^{3/2}i 2k_1.\xi_2k_{2d}\xi_{1a}(k_{1c}+k_{2c}+p_c) \Tr(P_{-}\fsC_{(n-1)}M_p\Gamma^{cda})L_1\label{56bbc7}\eeqa
once more  $k_{2c}$ has no contribution, taking  $pC=H$, we generate the third term ${\cal A}_{2}$  of  \reef{111}. One might suppose that $k_{1c}$ from \reef{56bbc7} is an extra singularity, however, we claim that the presence of this term is needed. Indeed if we take into account 
 the contribution $k_{1c}$ from \reef{56bbc7} and add it with the last term of ${\cal A'}_{2}$ we find
 
  \beqa
 2^{3/2}i 2k_1.\xi_2 k_{1c}\xi_{1a} (k_{2d}+k_{3d}) \Tr(P_{-}\fsC_{(n-1)}M_p\Gamma^{cda})L_1\label{56bbc}\eeqa
 which is exactly the 3rd term  ${\cal A}_{2}$. Therefore, we are able to produce not only all t-channel gauge field singularities \reef{111} but also its all  
 $s',u',(t+s'+u')$ channel tachyon singularities of $<V_{C^{-2}} V_{A^0} V_{A^0}V_{T^{0}}>$  are produced without any ambiguity. Let us deal with singularities that appear in \reef{113a}. Evidently ${\cal A''}_{3}$ is the same as ${\cal A}_{3}$. The 4th and 5th terms of ${\cal A''}_{2}$ are equivalent to 5th and 7th terms of ${\cal A}_{2}$. Applying momentum conservation to the 6th term  ${\cal A''}_{2}$, we would get
 \beqa
  2 L_1 k_{1d} \xi_1.\xi_2 \Tr(P_{-}\fsH_{(n)}M_p \Gamma^{cd}) (k_{1c}+k_{2c}+p_c)
\label{5673}\eeqa 
  $k_{1c}$ has no contribution to the  above interaction. The contribution from $k_{2c}$ produces exactly the 8th term ${\cal A}_{2}$, and to make a consistent result  for both symmetric amplitudes, one must impose the following  restricted Bianchi identity for the field strength of RR 
 
 \beqa
p_c H_{a_{0}...a_{p-2}} \epsilon^{a_{0}...a_{p-2}cd}=0\label{987}\eeqa

Using direct scattering amplitude $<V_{C^{-2}} V_{\phi^{0}} V_{\phi^0} V_{T^{0}}>$  the following Bianchi identity holds in terms of the both field strength of RR and RR potential  in the complete space-time 
 \beqa
 \eps^{a_0\cdots a_{p}} \bigg( -p_{a_{p}} (p+1) H^{ij}_{a_0\cdots a_{p-1}}-p^j H^{i}_{a_0\cdots a_{p}}+p^i H^{j}_{a_0\cdots a_{p}}\bigg)&=&d H^{p+2}=0.
 \label{BI12}\eeqa
 or
 \beqa
   p_{a_{0}} \eps^{a_0\cdots a_{p}} \bigg( -p_{a_{p}} p (p+1) C_{ija_1\cdots a_{p-1}}-p^j
   C_{ia_1\cdots a_{p}}+p^i C_{ja_1\cdots a_{p}}\bigg)&=&0.\label{mm11}\eeqa
Now if we again apply momentum conservation to the first and third terms  ${\cal A''}_{2}$ and simultaneously take into account the restricted world volume \reef{987}, then we actually reconstruct   the sum of  first and 6th terms of ${\cal A}_{2}$  as well as the 3rd and 4th terms  ${\cal A}_{2}$ accordingly. The same holds for the second term 
${\cal A''}_{2}$ and we regenerate the 2nd term ${\cal A}_{2}$. Hence, in comparison with \reef{111} and using the restricted world volume Bianchi identities 
we are able to produce  all t-channel gauge field singularities as well as
 $s',u',(t+s'+u')$ channel tachyon singularities of $<V_{C^{-1}} V_{A^0} V_{A^{-1}}V_{T^{0}}>$. It is worth emphasizing that unlike  $<V_{C^{-2}} V_{\phi^0} V_{\phi^{0}}V_{T^{0}}>$ analysis, here  we have no bulk singularity structures at all.  Hence,  on the world volume of non-BPS branes all brane singularities have been matched without producing any extra residual contact interactions.  
 
 \section{Contact term comparisons  }
To be able to achieve all the restricted Bianchi identities of non-BPS branes, we try to compare all contact interactions between \reef{44aa} and \reef{113a} with all order contact terms of \reef{111}. If we replace $k_{3c}=-(k_{1c}+k_{2c}+p_c)$ to  ${\cal A'}_{1}$ of \reef{44aa} (also to ${\cal A''}_{1}$) and use the following Bianchi identity 
\beqa
p_c \epsilon^{a_0..a_{p-5}cbda}=0\label{es99}\eeqa then we obviously produce all contact interactions of ${\cal A}_{1}$ for $p=n+3$ case. The leading order couplings can be produced if we would normalize the amplitude by $(\mu'_p\beta'\pi^{1/2})$ and compare it with the following coupling in the EFT
\beqa
\beta'\mu_p'(2\pi\alpha')^{3}\Tr (C_{p-4}\wedge F\wedge F\wedge DT)\labell{hderv}
\eeqa

All order contact interactions can be calculated. Recently the method of getting all order contact interactions has been released in \cite{Hatefi:2012wj,Hatefi1609}, so one can properly apply the higher derivative corrections to EFT couplings to produce all non-leading terms and also derive all order $\alpha'$ corrections. For example, if we replace the expansion of $L_3$ into the amplitude, then one can derive all order contact interactions of the amplitude for $p=n+3$ case  as follows
\beqa
8\beta'(\pi\alpha'^3)\mu_p'\bigg[\sum_{n=0}^{\infty}c_{n}\left(\frac{\alpha'}{2}\right)^{n}(D^aD_a)^n\Tr (C_{p-4}\wedge F\wedge F\wedge DT)\nonumber\\+\sum_{p,n,m=0}^{\infty}f_{p,n,m}\left(\frac{\alpha'}{2}\right)^{p}(D^aD_a)^p\left(\alpha'\right)^{2m+n}
C_{p-4}\wedge \Tr\bigg( D^{a_1}\cdots D^{a_{m}}  D^{b_1}\cdots D^{b_{n}} \nonumber\\ ((F\wedge  D^{a_{m+1}}\cdots D^{a_{2m}}F)\wedge D_{b_1}\cdots D_{b_{n}}D_{a_{1}}\cdots D_{a_{2m}}DT)\bigg)\bigg]\labell{hderv12}\eeqa

Note that  both ${\cal A''}_{1}$   and ${\cal A'}_{1}$ satisfy Ward identity associated to the gauge fields.

\vskip.1in
 
  Making use of the Bianchi identities we are able to generate  all contact interactions $<V_{C^{-2}} V_{A^0} V_{A^0}V_{T^{0}}>$  from \reef{111} without any ambiguity.  For instance, the first contact term of the amplitude for $p=n+1$ case is 
  \beqa
&&\frac{32}{(p-1)!}(\mu'_p\beta'\pi^{2})H_{a_{0}\cdots a_{p-2}}\xi_{1a_{p}}\xi_{2a_{p-1}}\eps^{a_{0}\cdots a_{p}}\label{uu67}\eeqa
and this contact interaction can be reconstructed by taking into account the following gauge invariant coupling in an EFT 
  \beqa
2\beta'\mu_p'(2\pi\alpha')^2\Tr(C_{p-2}\wedge F\wedge DT)\label{yy34}
\eeqa
Notice that  \reef{hderv} is found
by expanding the exponential of WZ action and using the multiplication rule of the super matrices.  
If we consider the expansions of $L_1,L_2$ into the amplitude then one may find the following contact interactions to the next leading order for $p=n+1$ case as below
\beqa
&&\frac{32}{(p-1)!}(\mu'_p\beta'\pi^{2})H_{a_{0}\cdots a_{p-2}}\eps^{a_{0}\cdots a_{p}} \bigg\{-\frac{\pi^2}{6}
\bigg(2k_2.\xi_1 k_{2a_{p-1}}\xi_{2a_{p}} -2k_1.\xi_2 k_{1a_{p-1}}\xi_{1a_{p}}\nonumber\\&&+2k_1.\xi_2\xi_{1a_{p-1}}k_{2a_{p}}+2k_2.\xi_1\xi_{2a_{p}}k_{1a_{p-1}}-t\xi_{1a_{p}}\xi_{2a_{p-1}}+2\xi_1.\xi_2k_{1a_{p}} k_{2a_{p-1}}\bigg)\bigg[t+2(s'+u')\bigg]\nonumber\\&&+\frac{\pi^2}{6}
\xi_{1a_{p}}\xi_{2a_{p-1}}(s'+u')^{2}+\bigg(\frac{\pi^2}{3}k_3.\xi_1k_{2a_{p-1}}\xi_{2a_{p}}\bigg[2(t+u')+s'\bigg]-[1\leftrightarrow 2]\bigg)\bigg\}
\labell{69aa}\eeqa
All terms in \reef{69aa} are related to the corrections of the EFT couplings. One can explore the following EFT couplings that  regenerate the contact terms in \reef{69aa} as follows
\beqa
-\frac{1}{12}\beta'\mu_p'(2\pi\alpha')^4\!\!\!\!\!\!\!\!&\bigg[-iD^{\beta}F_{a\alpha }D^{\alpha}F_{b\beta}D_cT+\frac{3i}{2}F_{ac}D_{\alpha}F_{\beta b}D^{\alpha}D^{\beta}T-\frac{3i}{2}D_{\alpha}F_{\beta b }F_{ac}D^{\alpha}D^{\beta}T\nonumber\\&
-\frac{1}{2}D_aD^{\alpha}D_cF_{b\alpha}D_{\beta}D^{\beta}T +F_{a\alpha }D^{\beta}D^{\alpha}D_{\beta}D_bD_cT
-\frac{1}{2}D_aD^{\alpha}D_{\beta}D^{\beta}F_{b\alpha}D_cT&\nonumber\\&+D_bD_cF_{a\alpha}D^{\beta} D^{\alpha}D_{\beta}T
+4D^{\alpha}D_a D_cF_{\beta b}D_{\alpha}D^{\beta}T-\frac{1}{2}D_aF_{\alpha \beta}D_bD^{\alpha}D^{\beta}D_cT&\nonumber\\
&-D_aD^{\beta}D_{\beta}D_cF_{b\alpha}D^{\alpha}T+2D_bD^{\alpha}D^{\beta}F_{a\alpha}D_{\beta}D_cT+D^{\alpha}D_{\alpha}D_cF_{\beta b }D^{\beta}D_aT&\nonumber\\
&+D_aD^{\beta}D_{\beta}F_{ b\alpha }D^{\alpha}D_cT+\frac{1}{2}D^{\beta}D^{\alpha}D_{\beta}D_cF_{a\alpha}D_b T&\nonumber\\
&-\frac{1}{2}D^{\alpha}D^{\beta}F_{ab} D_{\alpha}D_{\beta}D_cT\bigg]
\frac{1}{(p-2)!}C_{a_{0}\cdots a_{p-3}}\eps^{a_{0}\cdots a_{p-3}abc}&\labell{699bcc}
\eeqa
where the covariant derivative of tachyon is $D_aT=\prt_aT-i[A_a,T]$. Note that by direct scattering amplitude of a closed string RR, a tachyon and a gauge field in section two, we have derived all order $\alpha'$ higher derivative corrections to the last coupling of \reef{699bcc}. Let us now produce all $(t+s'+u')$-channel tachyon singularities of the amplitude.

\subsection{All  $(t+s'+u')$-channel tachyon singularities}
Let us  explore all $(t+s'+u')$- channel tachyon singularities of the amplitude ${\cal A'}_3$ for $p+1=n$ case. Extracting the trace and normalizing the amplitude we derive them as follows
 \beqa
&&p_c C_{a_{0}\cdots a_{p-1}}\eps^{a_{0}\cdots a_{p-1}c}\bigg(-t(k_3.\xi_1)(k_3.\xi_2)+(k_3.\xi_2)(k_2.\xi_1)s'+(k_3.\xi_1)(k_1.\xi_2)u'+\frac{1}{2}(\xi_1.\xi_2)s'u' \bigg)\nonumber\\&&\times
 \sum_{n,m=0}^{\infty}c_{n,m} (s'^{m}u'^{n}+s'^{n}u'^{m})  \frac{32 \beta'\mu'_p\pi^3}{(s'+t+u')p!}\label{88hh}\eeqa
which satisfies Ward identity. These poles can be constructed by employing a WZ coupling $ 2i\mu'_p\beta'(2\pi\alpha')\int C_{p}\wedge DT$ and all order higher derivative corrections to two tachyon-two gauge field couplings. In the effective field theory all singularities are derived by the following sub amplitude and vertices
\beqa
{\cal A}&=&V^{\alpha}(C_{p},T)G^{\alpha\beta}(T)V^{\beta}(T,T_3,A_1,A_2)\nonumber\\
G^{\alpha\beta}(T) &=&\frac{i\delta^{\alpha\beta}}{(2\pi\alpha') T_p
(s'+u'+t)}\nonumber\\
V^{\alpha}(C_{p},T)&=&2i\mu'_p\beta'(2\pi\alpha')\frac{1}{p!}\epsilon^{a_0\cdots a_{p}}C_{a_0\cdots a_{p-1}} k_{a_p}\labell{Fey}
\eeqa
Replacing the vertex of two tachyon-two gauge field couplings in the above field theory amplitude, we obtain all tachyon singularities in the EFT as follows
\beqa
&&32\pi\alpha'^{2}\beta'\mu_p'\frac{ \eps^{a_{0}\cdots a_{p-1}c} p_c C_{a_{0}\cdots a_{p-1}}}{p!(s'+t+u')}
\sum_{n,m=0}^{\infty}\bigg((a_{n,m}+b_{n,m})[s'^{m}u'^{n}+s'^{n}u'^{m}]\nonumber\\&&\times
\bigg[-t(k_3.\xi_2)(k_3.\xi_1)+(k_2.\xi_1)(k_3.\xi_2)s'+(k_1.\xi_2)(k_3.\xi_1)u'+(\xi_1.\xi_2)\frac{1}{2}u's'
\bigg]\bigg)\label{amphigh}\eeqa
Some of the coefficients are  
 \beqa
&&a_{0,0}=-\frac{\pi^2}{6},\,b_{0,0}=-\frac{\pi^2}{12}, a_{1,0}=2\z(3),\,a_{0,1}=0,\,b_{0,1}=b_{1,0}=-\z(3)\nonumber\eeqa
These poles \reef{amphigh} are exactly the ones that appeared in S-matrix elements \reef{88hh}.
  \subsubsection{All $u',s'$ channel-tachyon singularities }
Given the symmetries of the amplitude, we reconstruct all  $u'$-channel poles in the EFT, and likewise by exchanging momenta and polarizations,  all $s'$-channel singularities can also be examined.
\beqa
 \frac{32 \mu'_p\beta'\pi^{2}}{(p-2)!}  p_c C_{a_{0}\cdots a_{p-3}} 
\eps^{a_{0}\cdots a_{p-3}cae}\sum_{n=-1}^{\infty}b_n \frac{(s'+t)^{n+1}}{u'}(2k_3.\xi_2)k_{1a}\xi_{1e}\labell{Fey36}
\eeqa
 All these $u'$-channel poles can be constructed by the following rule
\beqa
{\cal A}&=&V^{\alpha}(C_{p-2},A_1,T)G^{\alpha\beta}(T)V^{\beta}(T,T_3,A_2)\labell{amp427}\eeqa
$V^{\beta}(T,T_3,A_2)$  should be found from the non-Abelian kinetic term of the tachyons in DBI action which has been fixed. If we employ the corrections that we got from WZ coupling $2i\beta'\mu'_p \int C_{p-2}\wedge F\wedge DT$ in \reef{highaa}, then we obtain the higher order vertex of 
$V^{\alpha}(C_{p-2},A_1,T)$  and the other vertices as follows
\beqa
V^{\beta}(T,T_3,A_2)&\!\!\!\!=\!\!\!\!&iT_p(2\pi\alpha')(k_3-k).\xi_2\nonumber\\
V^{\alpha}(C_{p-2},A_1,T)&\!\!\!\!=\!\!\!\!&2\mu'_p\beta'\frac{(2\pi\alpha')^{2}}{(p-2)!}\epsilon^{a_0\cdots a_{p-1}c} p_c C_{a_0\cdots a_{p-3}}k_{1a_{p-2}}\xi_{1a_{p-1}}\sum_{n=-1}^{\infty}b_n(\alpha'k_1\cdot k)^{n+1}
\label{66r}\eeqa
 $k$ is off-shell 's tachyon momentum. Replacing \reef{66r} inside \reef{amp427}, we obtain all order $u'$ channel tachyon poles  in an EFT as
\beqa
{\cal A}&=&\frac{2\mu'_p\beta'(2\pi\alpha')^{2}}{(p-2)!u'}\eps^{a_{0}\cdots a_{p-1}c} p_c C_{a_{0}\cdots a_{p-3}} k_{1a_{p-2}}\xi_{1a_{p-1}} (2k_3.\xi_2)\sum_{n=-1}^{\infty}b_n (t+s')^{n+1}\nonumber\eeqa
which are precisely  those singularities that appeared in  \reef{Fey36}. Eventually, one can show that all t-channel gauge field singularities are generated by taking into account the following rule and vertices in the EFT
\beqa
{\cal A}&=&V^{a}(C_{p-2},T_3,A)G^{ab}(A)V^{b}(A,A_1,A_2)\nonumber\\
V^{a}(C_{p-2},T_3,A)&=&2\mu'_p\beta'(2\pi\alpha')^{2}\frac{1}{(p-2)!} \eps^{a_{0}\cdots a_{p-2}ac} p_c C_{a_{0}\cdots a_{p-3}}k_{a_{p-2}}\sum_{n=-1}^{\infty}b_n(\alpha'k_3\cdot k)^{n+1}\nonumber\\
V^{b}(A,A_1,A_2)&=&-iT_p(2\pi\alpha')^{2} [\xi_{1}^{b}(k_1-k).\xi_{2}
+\xi_{2}^{b}(k-k_2).\xi_{1}+\xi_{1}.\xi_{2}(k_2-k_1)^{b}]\nonumber\\
G^{ab}(A)&=&\frac{i\delta^{ab}}{(2\pi\alpha')^{2}T_p t}\nonumber\eeqa
 $k$ is off-shell gauge field's momentum and $V^{a}(C_{p-2},T_3,A)$ has been derived from the corrections to the WZ coupling $C_{p-2}\wedge F\wedge DT$ that  we derived in the previous section. Notice that the kinetic term of the gauge fields is fixed in DBI action, so one reveals that $V^{b}(A,A_1,A_2)$ should not receive any  higher derivative corrections. We also come to know that the tachyon expansion that we talked about is consistent with effective field theory, because we are able to produce precisely all tachyon and massless poles of the string amplitude in the EFT as well.
 
 The expansion has also been checked for various other non supersymmetric cases, such as  all the other three and four point functions (like $CAT, C\phi \phi T$). That is why we believe that the expansion is universal. This might indicate that tachyon momentum expansion is unique. It would be nice to check it with the higher point functions of non-BPS string amplitudes. The precise form of the solutions for integrals of six point functions is unknown, however, given the exact symmetries of the string theory amplitudes and the universal tachyon expansion in  \cite{Hatefi:2017ags} we were able to obtain all the singularity structures of the amplitude of a closed string RR and four tachyons. We hope to overcome some other open questions in near future. 

\section*{Acknowledgements}

This paper was initiated during my 2nd post doc at Queen Mary University of London. Some parts of the paper were carried out at Mathematical institute in Charles University, at KITP in Santa Barbara, UC Berkeley and at Caltech. I am very grateful to  L. Alvarez-Gaume, K. Narain, F. Quevedo, D. Francia, A. Sagnotti, B. Jurco,  N.Arkani-Hamed, A. Brandhuber, G.Travaglini, P. Horava,  G.Veneziano, P. Sulkowski, P. Vasko, L. Mason, H. Steinacker and J.Schwarz  for  many useful discussions and for sharing their valuable insights with me. This work is supported by  ERC Starting Grant no. 335739
'Quantum fields and knot homologies', funded by the European Research Council.

  \end{document}